\documentclass[aps,pra,preprint,groupedaddress]{revtex4-1}
\usepackage{graphicx}
\usepackage{longtable}

\begin{document}

\title{Scattering of an alkali atomic beam on anti-spin-relaxation-coatings}

\author{Naota Sekiguchi}
\affiliation{Department of Applied Physics,
Tokyo University of Agriculture and Technology,
Koganei, Tokyo 184-8588, Japan}

\author{Kazane Okuma}
\affiliation{Department of Organic and Polymer Materials Chemistry,
Tokyo University of Agriculture and Technology,
Koganei, Tokyo 184-8588, Japan}

\author{Hiroaki Usui}
\affiliation{Department of Organic and Polymer Materials Chemistry,
Tokyo University of Agriculture and Technology,
Koganei, Tokyo 184-8588, Japan}

\author{Atsushi Hatakeyama}
\email{hatakeya@cc.tuat.ac.jp}
\affiliation{Department of Applied Physics,
Tokyo University of Agriculture and Technology,
Koganei, Tokyo 184-8588, Japan}

\date{\today}

\begin{abstract}
We performed scattering experiments using a rubidium (Rb) atomic beam on paraffin films and measured the angular and velocity distributions of scattered atoms. The paraffin films were prepared in various ways and characterized by atomic force microscopy and X-ray diffraction. The films exhibited various roughnesses and crystal structures.
The paraffin films preserved the spin polarization of the scattered atoms.
The measured angular distributions of all prepared films were consistent with Knudsen's cosine law.
The velocity distributions were well fitted by Maxwell's distribution, characterized by a temperature much closer to the film temperature than to the atomic-beam temperature.
We therefore concluded that the Rb atoms were well thermalized with the paraffin films via single scattering events.
\end{abstract}
\maketitle

\section{introduction%
\label{sec: introduction}%
}

Anti-spin-relaxation coatings on the inner walls of alkali vapor cells are used to preserve the spin polarization of alkali atoms in the cell \cite{Rob58}.
Anti-spin-relaxation-coated vapor cells have been applied to experiments requiring a long spin-relaxation time, such as frequency standards \cite{Rob82,Ban12}, ultra-sensitive magnetometry \cite{Dup69,Bud00,Bud07}, and quantum memory \cite{Jul04}.
Recently, novel experimental systems using a coated cell have been reported, e.g., an anti-PT symmetry optical experiment \cite{Pen16} and interferometry using a warm alkali-metal vapor \cite{Bie17}.
The behavior of atoms in a coated cell has attracted much attention from the research community. 

In the first decade after the discovery of paraffin as an anti-spin-relaxation coating material \cite{Rob58}, Bouchiat \textit{et al.} investigated the behavior of alkali atoms on the surface of paraffin \cite{Bou66} and proposed a mechanism to describe their interaction. 
Alkali atoms adsorb to the coating and remain there for some time
before undergoing desorption. 
The adsorption energy and dwell time on paraffin are 0.1~eV and on the order of a nanosecond, respectively.
Adsorption energy and dwell time are important parameters used to characterize the strength of the interactions between atoms and the coating; thus, there have been numerous related studies of alkali atoms on paraffin \cite{Gol61, Bre63, Lib86, Rah87, Ste94, Bud05, Ula11} and other coating materials, such as octadecyltrichlorosilane (OTS) \cite{Ste94, Yi08, Zha09, Ula11} and polydimethylsiloxane (PDMS) \cite{Atu99}.
Some studies have shown that alkali atoms diffuse into the coating \cite{Lib86, Bal95, Atu99, Bal12}.

The angular and velocity distributions of the desorbed atoms from the coating material also influence the behavior of atoms on the coating
\cite{Sti68}.
From a practical perspective, the angular and velocity distributions provide insight into the transport of atoms inside a confined device that has a coating \cite{Mat17}, given that the effect of atom-surface scattering on the atomic flow becomes more pronounced as the device becomes miniaturized.
Additionally, a better understanding of atom transport from the coating will be useful for laser cooling and trapping of short-lived radioactive alkali isotopes \cite{Sim96, Lu97, Atu03, Sak11} for electric dipole moment and parity-nonconservation interaction investigations.
However, the distributions have been inferred in only a few experiments \cite{Fru87,Kle11,Xia13}.
Some of the experimental results are in good agreement with theoretical predictions in which the angular distribution of the atoms obeys Knudsen's cosine law, a consequence of Maxwell-Boltzmann statistics \cite{Kle11,Xia13}.
On the other hand, 
non-Maxwellian distributions are required to explain the results of other experiments \cite{Fru87}.
Notably, 
the collisions of alkali atoms with background gas should be taken into account in coated cells, 
given that
the mean free path of an alkali atom in background gas 
as a result of chemical reactions with the coating 
has been estimated to be shorter than typical cell dimensions \cite{Sek16}.

One powerful and direct method used to investigate scattering behavior 
is scattering of an alkali atomic beam on a coating.
To date, 
scattering experiments involving alkali atomic beams on anti-spin-relaxation 
coatings
have not been reported.
There have only been a few scattering experiments of alkali atoms on metals or crystals \cite{Tay30, Ell37, McF60b, And86},
most of which showed that 
angular distribution followed Knudsen's cosine law 
and velocity distribution was characterized by a Maxwell distribution via the surface temperature
\cite{Tay30, Ell37, McF60b, Sti68, And86}.
In contrast, several experiments involving a LiF crystal \cite{McF60b} and a polished-glass surface \cite{And86} reported angular and velocity distributions that were non-Maxwellian.

Here, 
we report direct measurement of the angular and velocity distributions of rubidium (Rb) atoms scattered from paraffin films.
The morphologies of the prepared paraffin films were observed by atomic force microscopy (AFM); 
the roughness (Ra) of the films differed considerably.
The crystal structures were characterized by X-ray diffraction (XRD), which showed that
the molecular orientations depended on the film fabrication technique.
A Rb atomic beam was scattered by the paraffin films.
The anti-spin-relaxation performance of the films was investigated 
by comparing the spin polarizations of the atomic beam and scattered atoms.
The angular and velocity distributions of scattered atoms were examined by detecting laser-induced fluorescence from the atoms.
The measured angular distributions of all films were well described by the cosine law. 
The velocity distributions were well fitted by the Maxwell velocity distribution and were characterized by temperatures much closer to the film temperature than to the atomic-beam temperature.
From these results, we concluded that
incident Rb atoms were well accommodated thermally by the paraffin film surface by single collisions, 
and spin polarization was preserved.

\section{apparatus%
\label{sec: apparatus}%
}
\begin{figure}[t]
	\centering
	\includegraphics[clip,width=.4\textwidth]{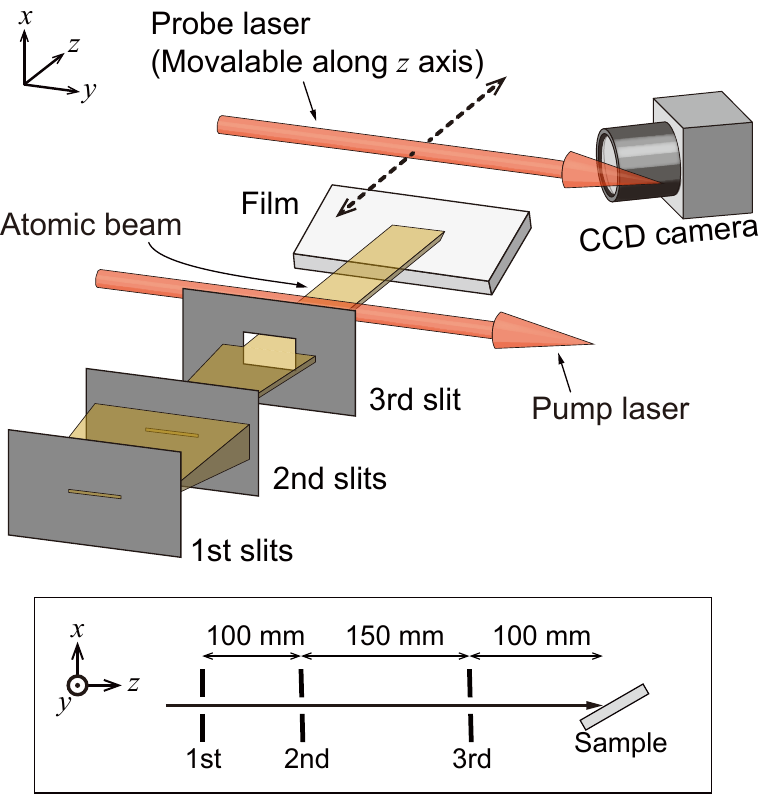}
	\caption{(Color online) Schematic diagram of the experiment. 
The rubidium (Rb) atomic beam was collimated by three slits 
	before colliding with the film.
	Scattered atoms were detected using probe laser light and a charge-coupled device (CCD) camera.
	The probe light moved along the $z$ axis.
	The incident atomic beam was spin-polarized by 
	the pump laser light.
	The inset drawing shows the 
	distances between the slits and the film.
	\label{fig: setup}}
\end{figure}

Figure~\ref{fig: setup} shows a conceptional sketch of our experiment.
The Rb atomic beam emerged from an oven 
and was collimated using three slits (1st, 2nd, and 3rd slits in Fig.~\ref{fig: setup}).
The collimated atomic beam collided
with the film mounted on a rotational and translational stage.
Atoms scattering from the film were illuminated with the probe laser light (diameter: 1.0~mm).
The fluorescence induced by the probe light was collected by a charge-coupled device (CCD) camera that was in the $y$-$z$ plane but not perpendicular to the $z$ axis.
The probe light position moved in the $z$ direction 
during exposure of the CCD camera to the fluorescence.
For spin polarization measurements, 
the pump light (diameter: 1.5~mm) was introduced upstream of the film.
The incident atomic beam was spin-polarized using the pump light, and the
light polarization was linear.
A beam shutter (not shown in Fig.~\ref{fig: setup}), mounted on a 
translational stage, 
was used to block the Rb atomic beam, 
enabling evaluation of the background signal.

The oven, slits, and film resided in a vacuum chamber 
maintained at a pressure of a few $10^{-5}$~Pa.
The oven temperature was maintained at 200$^{\circ}$C during measurements,
and the film was held at room temperature.
The three slits were rectangular; the first and second were
$0.1 \times 3$~mm$^{2}$ and the third was $2 \times 3$~mm$^{2}$.
The separations between the first and second, second and third slits, 
and the third slit and the film were 100~mm, 150~mm, and 100~mm, 
respectively, as shown in the inset of Fig.~\ref{fig: setup}.
As a consequence of the collimation, 
the angular spread of the Rb atomic beam was 1~mrad along the $x$ axis 
and 12~mrad along the $y$ axis.
The flux of atoms colliding with the film was estimated to be $8\times 10^{10}~\mathrm{s^{-1}}$ based on the oven temperature and slit geometries.

\begin{figure}[t]
	\centering
	\includegraphics[clip,width=.5\textwidth]{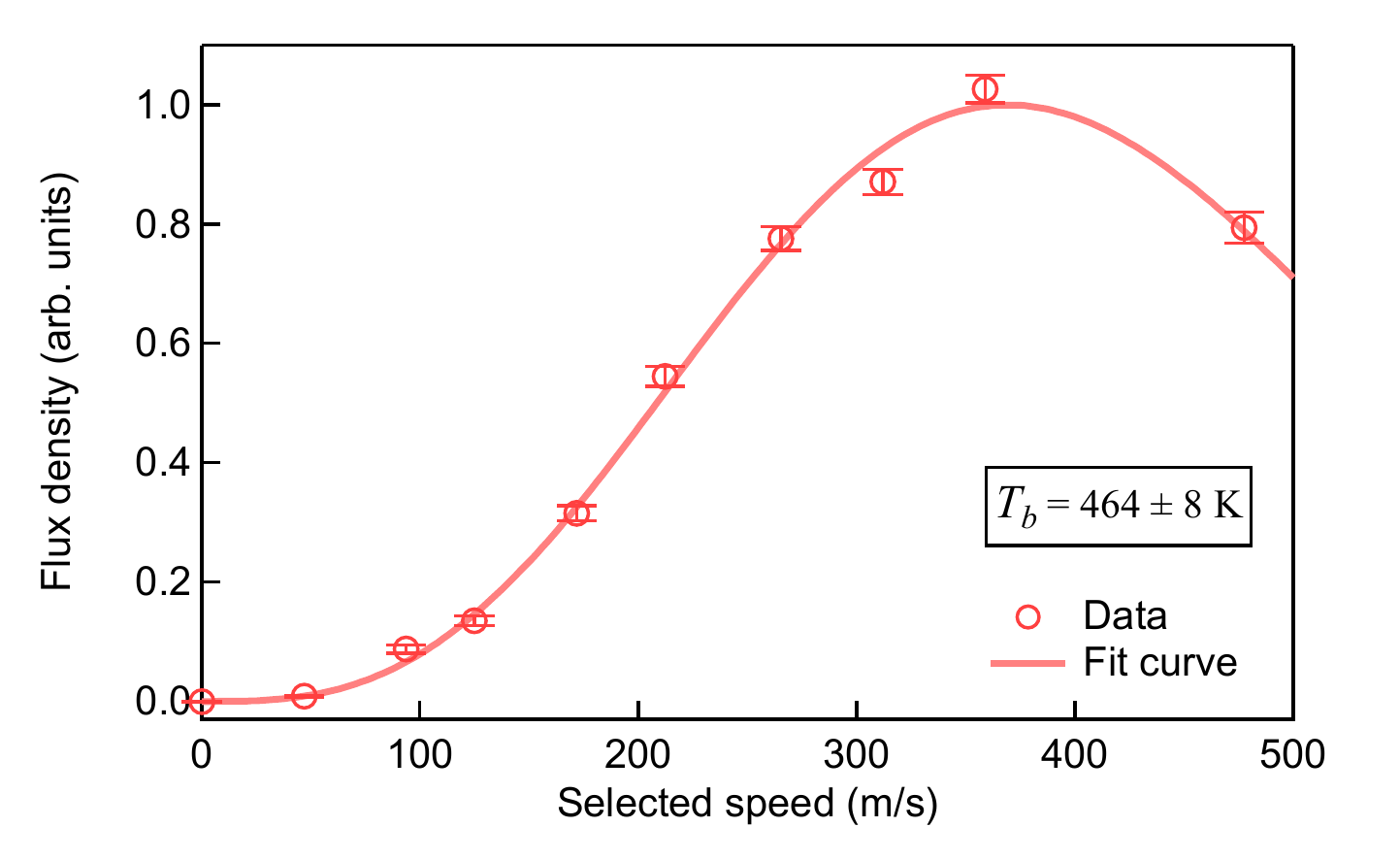}
	\caption{(Color online) Flux density $q_{b}(v)$ of the atomic beam as a function of the speed $v$ of atoms.
	Data (open circle) were fitted by the fit curve (solid line) given by Eq.~(\ref{eq: sd_beam}).
	The temperature of the atomic beam was estimated to be $464 \pm 8$~K.
	\label{fig: beamvd}}
\end{figure}

The temperature of the $^{85}$Rb atoms in the atomic beam 
was measured spectroscopically.
For temperature measurements, the film was moved out of the path of the atomic beam, 
and a second laser beam, counter-propagative to the atomic beam, was introduced to the vacuum chamber.
The frequency of the laser beam was red-detuned from the resonance frequency of a $^{85}$Rb atom at rest.
Laser-induced fluorescence from the atoms with a speed $v$ corresponding to the detuning was observed due to the Doppler effect.
The intensity of the fluorescence, $I_{b}$, was proportional to the number density $n_{b}(v)$ of atoms having speed $v$ in the atomic beam.
Also, 
the flux density $q_{b}(v)$ of the atoms with speed $v$ was proportional to the product of the fluorescence intensity and speed $v$,
because the flux density $q_{b}(v)$ was obtained by multiplying the number density $n_{b}(v)$ by the atomic speed $v$,
\begin{equation}
q_{b}(v)=vn_{b}(v) \propto v I_{b}.
\label{eq: fluxpropI}
\end{equation}
Figure~\ref{fig: beamvd} shows the flux $q_{b}(v)$ (open circle) as a function of the speed $v$ of atoms.
The vertical axis is normalized to show a unit value at the peak.
The uncertainty of measurements was estimated based on multiple measurements at a certain velocity;
the estimated standard deviations are represented by error bars.
The speed distribution $f_{b}(v)$ of an atomic beam flux is given by
\begin{equation}
f_{b}(v)=\frac{m^{2}}{2k_{B}^{2}T_{b}^{2}}v^{3}\exp \left( -\displaystyle \frac{mv^{2}}{2k_{B}T_{b}} \right),
\label{eq: sd_beam}
\end{equation}
where $m$ is the mass of a $^{85}$Rb atom, $k_{B}$ is Boltzmann's constant, and $T_{b}$ is the temperature of the atomic beam.
By fitting $f_{b}(v)$ with a scaling factor to the data $q_{b}(v)$, as shown by the solid curve in Fig.~\ref{fig: beamvd},
the temperature $T_{b}$ was determined to be $T_{b} = 464\pm 8$~K, which is consistent with the oven temperature. 

\section{Film preparation and characterization%
\label{sec: film}}
\begin{table*}
	\caption{\label{tb: films}List of films. 
	Film preparation methods, substrates, 
	arithmetic average of the roughness (Ra) within a field of $5\times 5~\mu \mathrm{m^{2}}$ after scattering experiments, 
	and crystalline characteristics after scattering experiments are shown.}
	\begin{ruledtabular}
		\begin{tabular}{cccccc}
				&	Method					&	Film							&	Substrate						&	Ra (nm)	&	Molecular orientation				\\	\hline
		\#1	&	---							&	$\mathrm{SiO_{2}}$	&	Si 									&	---				&	---											\\
		\#2	&	Dip coating			&	Tetracontane				&	$\mathrm{Si/SiO_{2}}$	&	2.6			&	Normal 									\\
		\#3	&	Vapor deposition	&	Tetracontane				&	Borosilicate glass 			&	0.9			&	Mainly random with some normal and lateral	\\
		\#4	&	Vapor deposition	&	Tetracontane				&	APS monolayer				&	0.7			&	Mainly random with some lateral					\\
		\end{tabular}	
	\end{ruledtabular}
\end{table*}

\begin{figure}[b]
	\centering
	\includegraphics[clip, width=0.45\textwidth]{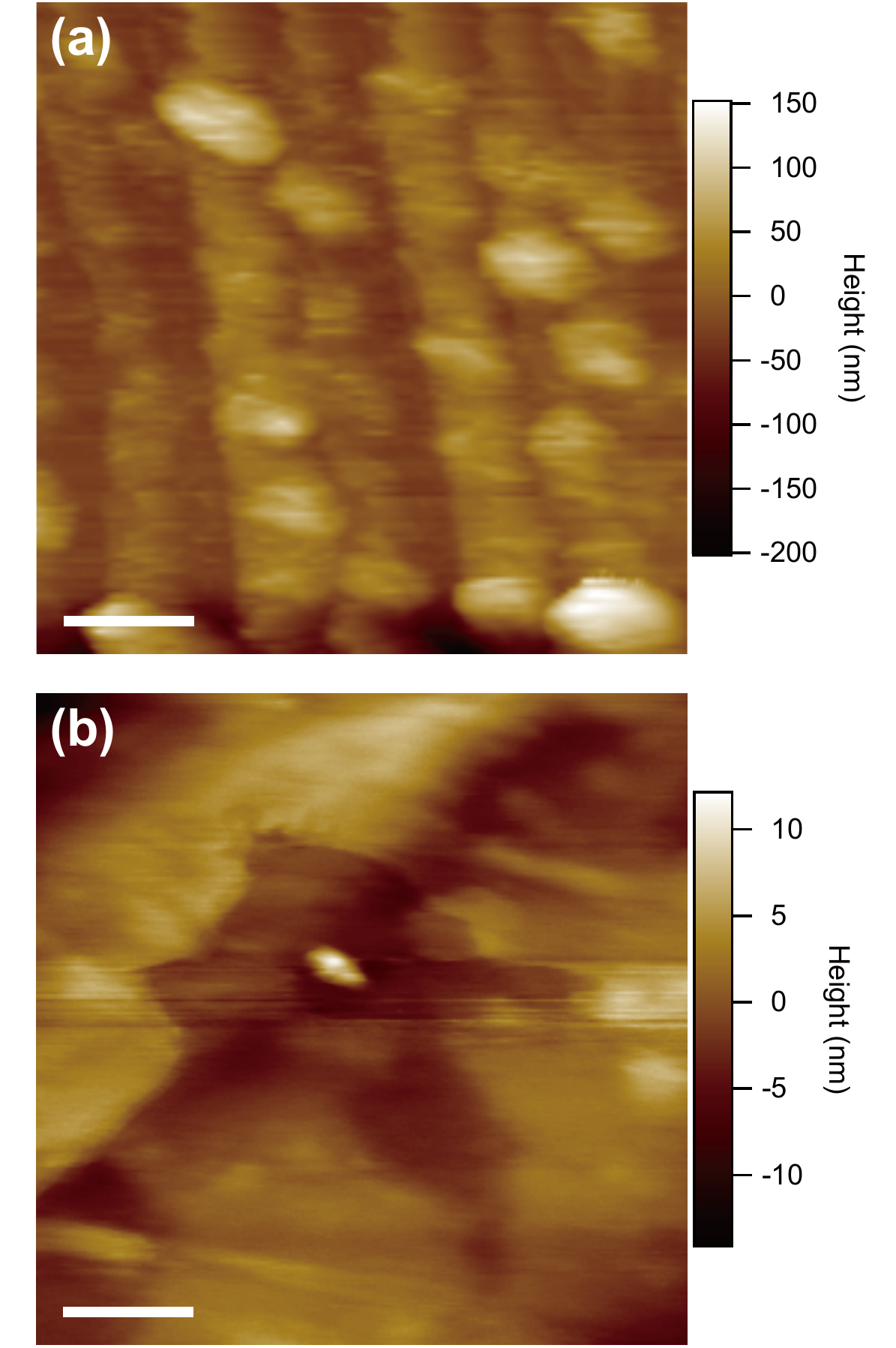}
	\caption{(Color online) 
	Atomic force microscopy (AFM) height images of film \#2 (a) before scattering experiments (Ra = 24~nm) 
	and (b) after scattering experiments (Ra = 2.6~nm).
	Note the different height scales 
	and AFM viewpoints.
	The image size was $5 \times 5~\mu\mathrm{m^{2}}$, 
	and the white bars indicate 1~$\mu$m.
	Heights of 0~nm represent the average height of the images.
	\label{fig: AFM2}}
\end{figure}

\begin{figure}[b]
	\centering
	\includegraphics[clip, width=0.5\textwidth]{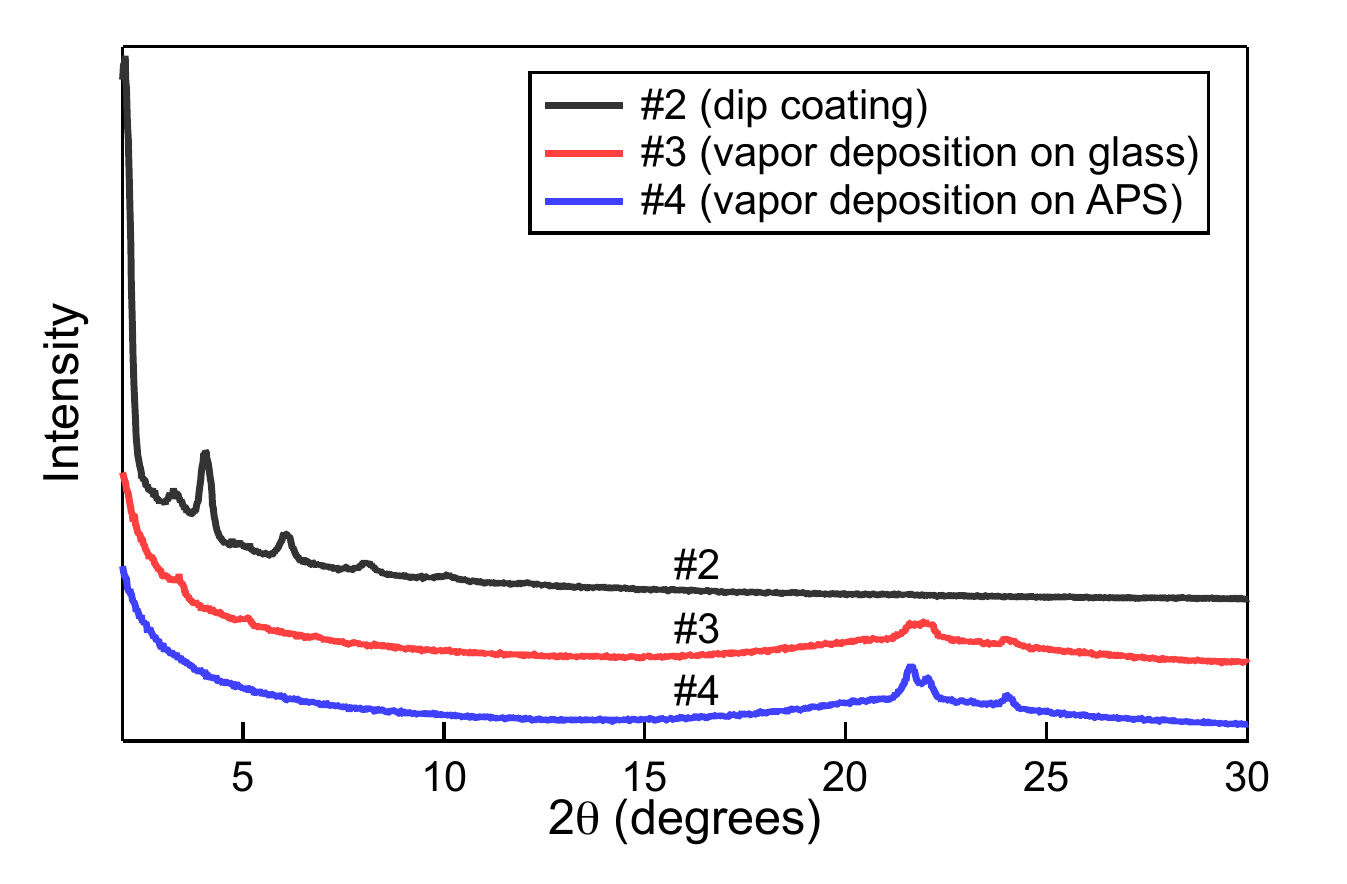}
\caption{(Color online) X-ray diffraction (XRD) spectra of tetracontane films. The spectra are offset for easier viewing.}
	\label{fig: XRD}
\end{figure}

The films examined in this study are summarized in Table~\ref{tb: films}.
Film \#1 was a bare $\mathrm{Si/SiO_{2}}$ plate
for comparison, 
and the other films were tetracontane 
($\mathrm{C_{40}H_{82}}$, Sigma-Aldrich, $>95.0$\% purity) coated onto substrates.

Tetracontane film \#2 was prepared on 
a $\mathrm{Si/SiO_{2}}$ plate using a dip coating method.
Dip coating is commonly used to produce flat, homogeneous films on 
substrates.
A very smooth thin film of tetracontane on a silica substrate can be produced by dip coating \cite{Hib12}.
Using an approach similar to that of Ref.~\onlinecite{Hib12}, 
we coated the $\mathrm{Si/SiO_{2}}$ substrate with tetracontane.
The silica substrate was cleaned with piranha solution for 10 min in 
a mixture of 30\% hydrogen peroxide ($\mathrm{H_{2}O_{2}}$)
and 96\% sulfuric acid ($\mathrm{H_{2}SO_{4}}$) with a volume ratio of $1:3$.
The cleaned substrate was rinsed with deionized water several times 
and dried under a flow of nitrogen gas.
During the dip-coating process, 
the substrate and tetracontane in a glass container were placed in an oven at 120$^{\circ}$C.
The substrate was dipped into the melted tetracontane and withdrawn at a constant speed.
After withdrawal, the oven was cooled slowly.
The thickness of film \#2 was a few hundred nanometers.

Film \#3 was a tetracontane thin film coated onto a borosilicate-glass substrate 
by vapor deposition.
The glass substrate was washed with detergent 
and ultrasonically cleaned with deionized water, acetone, ethanol, and methanol.
Then, the substrate was dried under nitrogen gas.
Tetracontane was evaporated at 300$^{\circ}$C and deposited onto the substrate kept at 30-45$^{\circ}$C 
for 45 min in a vacuum chamber.
The thickness of film \#3 was 107~nm.

Film \#4 was also prepared by vapor deposition of tetracontane 
but onto a self-assembled monolayer (SAM) of 3-aminopropyltrimethoxysilane (APS) 
on borosilicate-glass.
In addition to the cleaning processes for film \#3, 
the borosilicate-glass substrate was treated by ultraviolet and ozone exposure.
The substrate was then immersed in toluene with 1 wt\% APS for 1 h.
After immersion, 
the substrate was ultrasonically cleaned with toluene for 5 min 
and dried under a nitrogen atmosphere at 100$^{\circ}$C for 1 h.
APS molecules formed a SAM on the glass substrate using this procedure.
Vapor deposition of tetracontane onto the APS-SAM was performed using
the same procedure as that used for film \#3.
The thickness of film \#4 was 230~nm.

The tetracontane films were characterized by AFM 
and 
XRD analyses.
The surface morphologies of film \#2 before and after the scattering experiments 
were analyzed by AFM.
Figure~\ref{fig: AFM2} shows height images of film \#2 within a field $5 \times 5~\mu\mathrm{m^{2}}$.
The horizontal direction corresponds to the $y$ axis in Fig.~\ref{fig: setup}.
The color scales show the height with respect to the average height over the viewing area. The images show different viewpoints, revealing 
various modifications of the surface morphologies.
The modified surfaces were attributed to the incident atomic beam during the scattering experiments.
The arithmetic average of the Ra was evaluated from the heights in the images.
The Ra value of film \#2 decreased from 24~nm (before the scattering experiments) to 2.6~nm (after the scattering experiments).
The surface morphologies of films \#3 and \#4 after the scattering experiments
had 
Ra values of 0.9~nm and 0.7~nm for films \#3 and \#4, respectively.

Figure~\ref{fig: XRD} shows XRD spectra of tetracontane films after the scattering experiments.
The horizontal axis represents the diffraction angle $2\theta$, defined as the angle between incident and diffracted X-rays.
The vertical axis shows the intensity of the diffracted X-rays; spectra are offset vertically for easier viewing.
The diffraction peaks at a low diffraction angle, $2\theta \leq 10^{\circ}$, 
indicate normal molecular orientations \cite{Tan90}, 
and the peaks in the range 20$^{\circ}$ to 25$^{\circ}$ 
indicate lateral molecular orientations \cite{Tan90}.
The broad pedestal centered around 21$^{\circ}$ was attributed to 
the structure of the borosilicate-glass substrates.
Our results show that the tetracontane thin film obtained by dip coating (\#2) 
was assembled mainly with normal molecular orientations.
In contrast, 
the films grown by vapor deposition (\#3 and \#4) were 
composed mainly of randomly oriented molecules, 
because the spectra had small diffraction peaks.
Nonetheless, 
film \#3 had crystallites with normal and lateral orientations, 
and film \#4 had crystallites with lateral orientations. 

\section{Experiments%
\label{sec: experiments}%
}

\subsection{Angular distribution in the $x$-$z$ plane}

\begin{figure}[t]
	\centering
	\includegraphics[clip, width=0.4\textwidth]{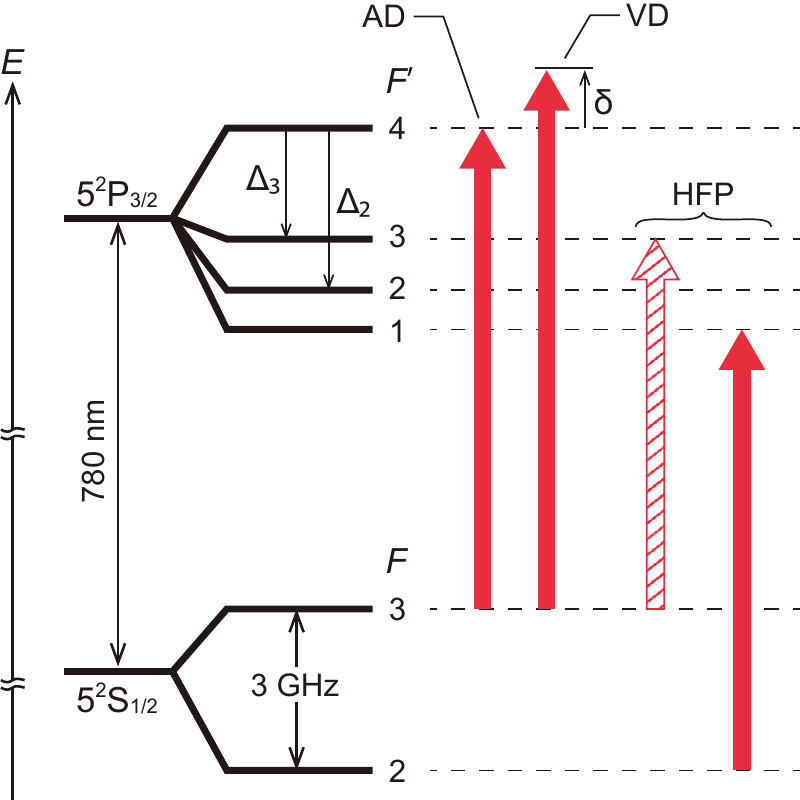}
	\caption{(Color online) 
	Energy level diagram of $^{85}$Rb (energy separations not to scale).
	Energy differences for $F'=3$ and $2$ from $F'=4$ are denoted by $\Delta_{3}(=121$~MHz) and $\Delta_{2} (=184$~MHz), respectively.
	The arrows show the frequencies of the probe laser (filled arrows) 
	and the pump laser (hatched arrow) used in each measurement, 
	angular distribution (AD) measurement, velocity distribution (VD) measurement, and 
	hyperfine polarization (HFP) measurement.
	The detuning $\delta$ of the probe laser for VD measurement can be up to $500$~MHz.
	\label{fig: energylevels}}
\end{figure}

We measured the angular distributions of scattered atoms in the $x$-$z$ plane.
The pump light was not used in these measurements.
The Rb atomic beam entered the film at an incident angle $\theta_{i}$ 
defined as the angle from the surface normal.
In this study, we fixed $\theta_{i} \simeq 70^{\circ}$.
Scattered atoms were irradiated by the probe light,
which was resonant with the transition $F=3 \to F'=4$ in the $D_{2}$ lines of $^{85}$Rb
(see Fig.~\ref{fig: energylevels}).
The absorption of the resonant probe light 
led to fluorescence emission from the scattered atoms.
Due to velocity selection along the $y$ axis (the laser direction) of around 0~m/s 
by the Doppler effect, 
we examined the atoms in the $x$-$z$ plane.
The CCD camera was exposed to the fluorescence for a certain period, 
while the position of the probe light was scanned along the $z$ axis.
The fluorescence images from different positions along $z$ were acquired. 

\begin{figure}[t]
	\centering
	\includegraphics[clip, width=0.45\textwidth]{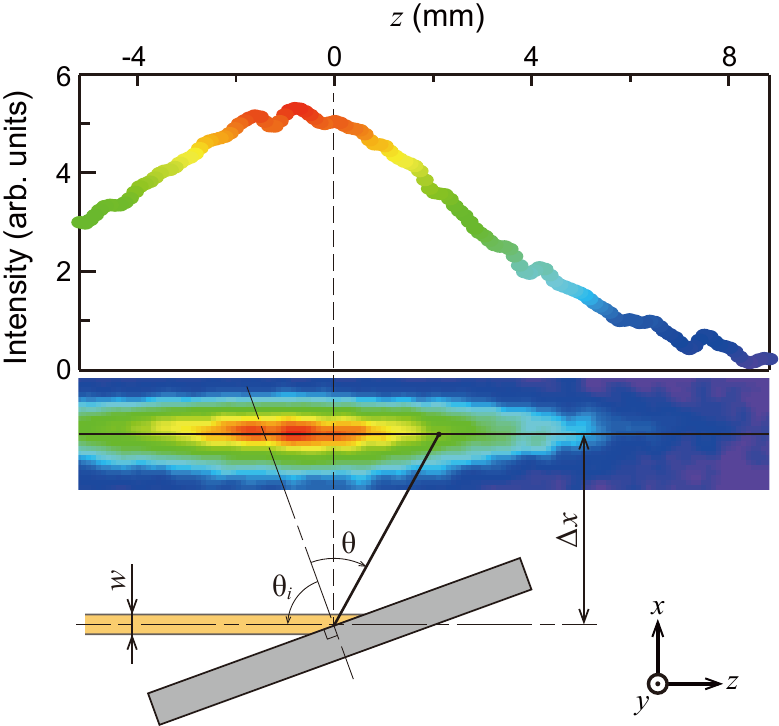}
	\caption{(Color online) 
	Fluorescence image in the angular distribution measurement.
	The geometry of the scattering experiments is illustrated to scale.
	The solid line in the image indicates the center line of the fluorescence image.
	The distance of the center line from the atomic beam is denoted by $\Delta x$.
	The upper graph shows the fluorescence intensity $I(z)$ on the center line.
	When this image was taken, 
	the CCD camera was exposed for 228~s, 
	and the probe light power was 200~$\mu$W.
	The other parameters were as follows: 
	$\theta_{i} = 70^{\circ}$; $w= 0.5$~mm; $\Delta x =3.9$~mm.
	\label{fig: LIF}}
\end{figure}

Figure~\ref{fig: LIF} shows an example of a fluorescence image.
For the image in Fig.~\ref{fig: LIF}, 
fluorescence from scattered atoms was induced by the probe light of 200~$\mu$W 
and recorded with the CCD camera for 228~s.
The width $w$ of the atomic beam along the $x$ axis was about $0.5$~mm.
The distance $\Delta x$ from the scattering point to the center line of the fluorescence 
was 3.9~mm.
Fluorescence from the background atoms and stray light 
were eliminated by subtracting the background image taken with the atomic-beam shutter.
The solid curve in the upper graph shows the dependence of the fluorescence intensity $I(z)$ on the position of $z$ along the center line of the image.
The fluorescence intensity value is indicated by the vertical axis and the curve's color, the color scale of which is the same as that of the fluorescence image.
The flux density $q$ of scattered atoms within the scattering angle 
$\theta$ is dependent on the position $(r, \theta)$, in polar coordinates, that is, $q=q(r,\theta)$.
Here, the scattering angle $\theta$ and the distance $r$ from the scattering point are expressed by
\begin{equation}
\theta = \tan^{-1}\left( \frac{z}{\Delta x} \right) + \left(\frac{\pi}{2}-\theta_{i}\right),
\end{equation}
and 
\begin{equation}
r=\sqrt{\Delta x^{2} + z^{2}},
\end{equation}
respectively.
Given the angular distribution $s(\theta)$ of scattered atoms, the flux density $q(r,\theta)$ can be expressed by 
\begin{equation}
	q(r,\theta) = \frac{s(\theta)d\theta}{rd\theta},
\end{equation}
where the numerator $s(\theta)d\theta$ represents the flux of atoms scattered within the angular range of $\theta$ to $\theta + d\theta$, and the denominator $rd\theta$ represents the arc length.
Similar to Eq.~(\ref{eq: fluxpropI}), 
the fluorescence intensity $I(z)$ is proportional to the flux density $q(r,\theta)$,
\begin{equation}
	I(z) \propto \frac{q(r,\theta)}{\bar{v}(\theta)} = \frac{s(\theta)}{\bar{v}(\theta)r}.
	\label{eq: Iprops}
\end{equation}
Here, $\bar{v}(\theta)$ is the mean speed of atoms at angle $\theta$.
The fluorescence intensity $I(z)$ therefore indicates the angular distribution $s(\theta)$.

\subsection{Velocity distribution along the $y$ axis}
In this measurement, 
the position of the probe light was not scanned but fixed at $\theta \sim 0^{\circ}$, 
and the pump light was not used.
The power of the probe light was 100~$\mu$W.
The frequency of the probe light was blue-detuned using an acousto-optic modulator (AOM) 
by an amount of $\delta$ from the transition frequency $F=3 \to F'=4$, as shown in Fig.~\ref{fig: energylevels}.
Due to the Doppler effect, 
scattered atoms moving at velocity $v_{y}$ along the $y$ axis were selectively detected by the detuned probe light and the CCD camera.
The velocity $v_{y}$ corresponded to the difference between the frequency of the detuned probe light and the transition to the $F'=2,3,\mathrm{or}~4$ states:
\begin{equation}
	v_{y}=\lambda_{0}(\delta-\Delta_{F'}),
	\label{eq: sv}
\end{equation}
where $\lambda_{0}$ is the wavelength of the $D_{2}$ line and 
$\Delta_{F'}$ is the splitting of the excited states from the $F'=4$ state in frequency.
The transition to $F'=4$ has the largest absorption cross-section 
among the transitions from the ground state $F=3$.
Furthermore, 
the excitations to the $F'=2$ and $3$ states depletes the population in the $F=3$ state, 
leading to less absorption of the probe light.
We therefore considered only the transition $F=3 \to F'=4$ 
in this measurement.
The measurements were repeated with different detuning frequencies $\delta$.
The intensities of the fluorescence as a function of the detuning $\delta$ 
reflect the velocity distribution of the scattered atoms along the $y$ axis.

\subsection{Hyperfine polarization}
We examined the anti-spin-relaxation performance of the films for the incident atoms.
Pump light of 50~$\mu$W illuminated the atomic beam in the upper 
stream 
of the film, as shown in Fig.~\ref{fig: setup}.
The frequency of the pump light was stabilized to the transition $F=3 \to F'=3$ of the $D_{2}$ line, 
as shown in Fig.~\ref{fig: energylevels} by the hatched arrow.
The pump light selectively excited atoms that had velocity $v_{y}$ along the $y$ axis around 0~m/s
within the velocity width of $\sim 5$~m/s corresponding to the natural line width of the transition.
The velocity selection width was comparable 
to the velocity width of the atomic beam along the $y$ axis 
estimated from the angular spread of 12~mrad and the mean speed of $\sim 500$~m/s.
Consequently, hyperfine polarization of the atoms was produced in every velocity group in the atomic beam, 
that is, 
the populations of the ground states 
were polarized to the $F=2$ hyperfine state between the ground states $F=2$ and $3$.
The probe light of 15~$\mu$W was tuned to the transition $F=2 \to F'=1$ of the $D_{2}$ line 
to probe the population in the $F=2$ state.
The fluorescence $I_{p}$ was induced by the probe light and recorded by the CCD camera.
We defined and evaluated the fluorescence difference,
\begin{equation}
	\Delta_{S} = \frac{I_{p}-I_{0}}{I_{0}},
\end{equation}
with $I_{0}$ as the fluorescence recorded in the absence of hyperfine pumping.
$\Delta_{S}$ indicates the difference in the population in the $F=2$ state from that in the non-polarized state (thermal equilibrium), 
given that the fluorescence intensity is proportional to the population in the $F=2$ state.
The difference $\Delta_{B}$ was also evaluated for the incident atomic beam in the same way.
For the non-polarized state, 
the population in the $F=2$ state is given by $g_{2}/(g_{2}+g_{3}) = 5/12$, where $g_{2}$ and $g_{3}$ are the number of sublevels in the $F=2$ and $F=3$ states, respectively.
With some algebra, the ratio $P=\Delta_{S}/\Delta_{B}$ was derived to be equal to the ratio of the differences in population between the ground states:
\begin{equation}
	P=\frac{\Delta_{S}}{\Delta_{B}}=\frac{g_{2}N_{3}^{S}-g_{3}N_{2}^{S}}{g_{2}N_{3}^{B}-g_{3}N_{2}^{B}}.
\end{equation}
Here, $N_{F}$ is the population in the ground state specified by $F$ when the atomic beam was hyperfine-polarized, and the superscripts $S$ and $B$ represent the values for the atomic beam and scattered atoms, respectively.
In this study, we measured the ratio $P$, the surviving hyperfine polarization, for all prepared films.

\section{Results and Discussions%
\label{sec: results}%
}

Before the measurements, 
the films were exposed to the atomic beam until the fluorescence from scattered atoms stabilized.
We observed that 
the $\mathrm{Si/SiO_{2}}$ (film \#1) required exposure for several hours 
before the scattering intensity stabilized, 
whereas the paraffin films were able to scatter atoms shortly after 
exposure.

\begin{figure}[t]
	\centering
	\includegraphics[clip, width=0.5\textwidth]{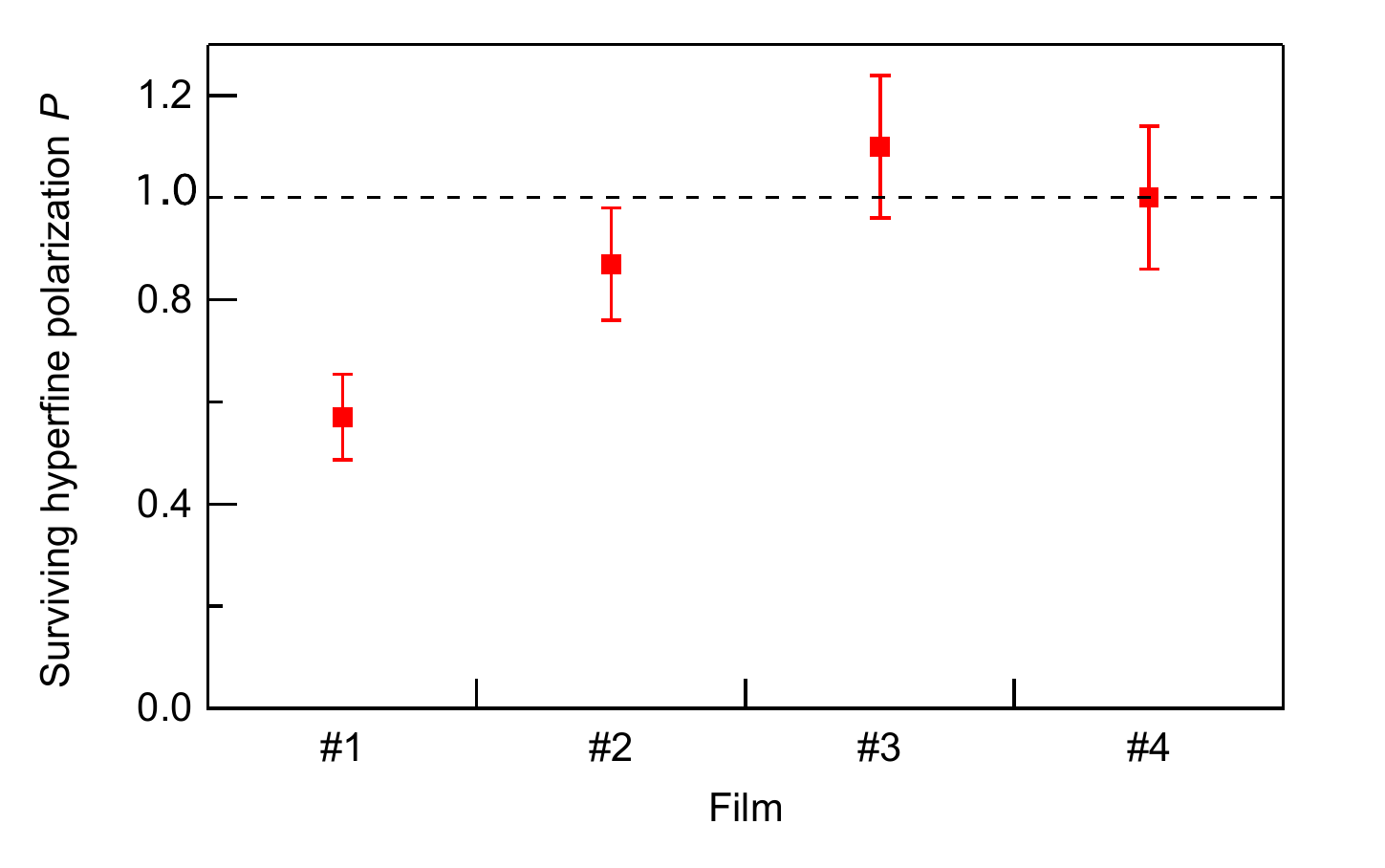}
	\caption{\label{fig: hf}
	Surviving hyperfine polarizations of atoms scattered from the films.
	The vertical axis is normalized by the hyperfine polarization of the incident atomic beam.}
\end{figure}

The surviving hyperfine polarizations $P$ for films \#1-4 
are shown in Fig.~\ref{fig: hf}.
The dashed line represents the unit value in $P$ and indicates no depolarization by scattering on the films.
We confirmed that
tetracontane films (\#2-4) preserved polarization during scattering.
It is interesting to note that the uncoated $\mathrm{Si/SiO_{2}}$ plate (film \#1) preserved 
half of the polarization of incident atoms by a single collision.

\begin{figure}[b]
	\includegraphics[clip, width=0.5\textwidth]{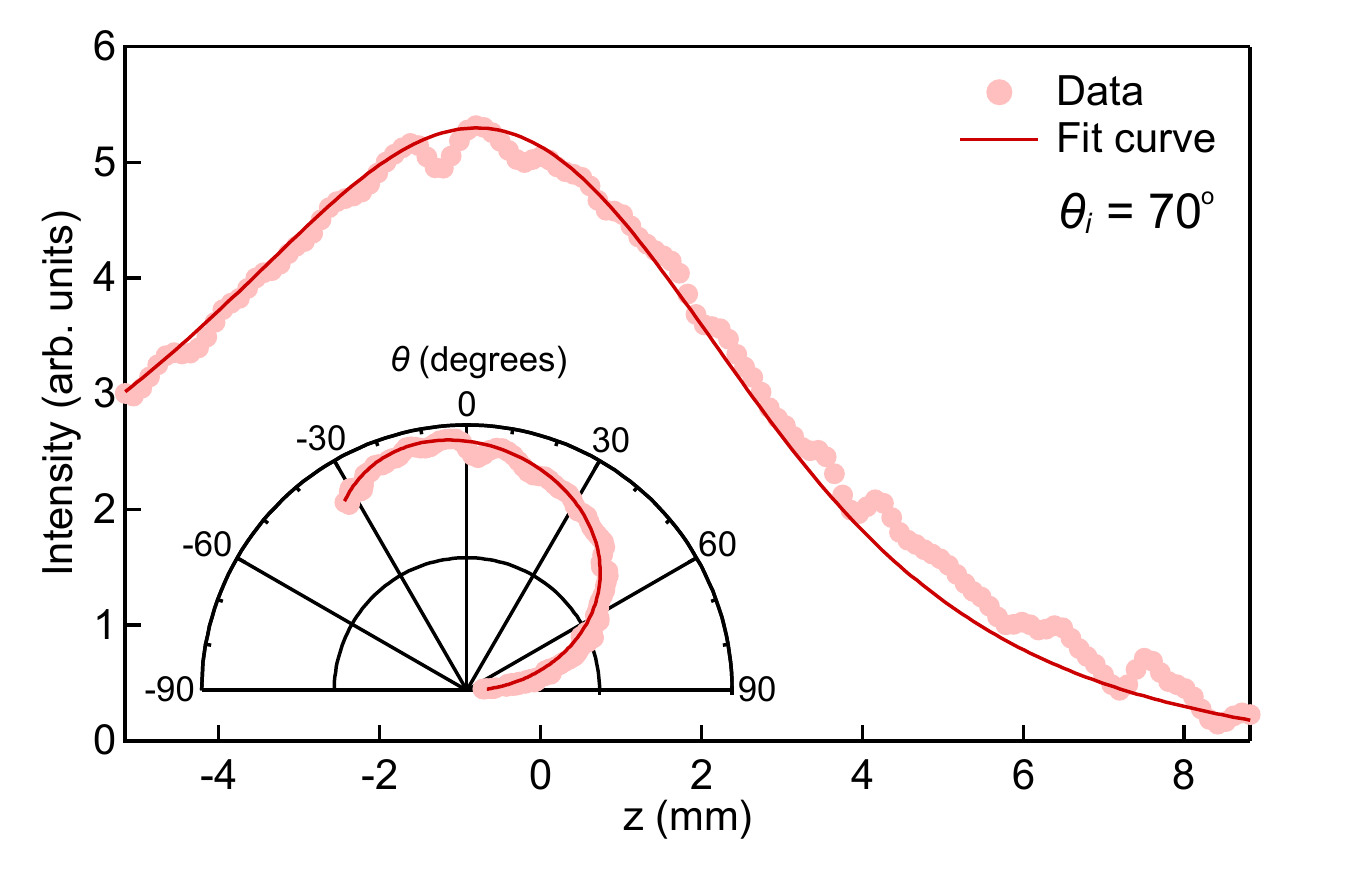}
	\caption{\label{fig: sp}
	(Color online) Fluorescence intensity $I(z)$ as a function of $z$ for film \#2 at the incident angle $\theta_{i}$ of 70$^{\circ}$. The points show the experimental data, and the solid curve is the fitted curve based on the cosine law. The inset polar graph shows the angular distribution $s(\theta)$ derived from $I(z)$.
	}
\end{figure}

Figure \ref{fig: sp} illustrates the fluorescence intensity $I(z)$ as a function of $z$, as shown in the upper graph in Fig.~\ref{fig: LIF}.
The points represent the experimental data, and 
the solid curve is the theoretical curve for an angular distribution that obeys the cosine law, $s(\theta) \propto \cos \theta$ and $\bar{v}(\theta)$ being independent of $\theta$, 
taking into account the experimental conditions, including the widths of the atomic beam and the shooting angle of the CCD camera.
The curve was fitted to experimental data with a scaling factor, which was the only fitting parameter applied.
From the figure, 
the data can be described by the cosine law.
The polar plot as a function of the scattering angle $\theta$ in the inset of Fig.~\ref{fig: sp} shows the angular distribution $s(\theta)$.
In the derivation of the angular distribution $s(\theta)$ from $I(z)$ using Eq.~(\ref{eq: Iprops}), the mean speed $\bar{v}(\theta)$ was considered independent from $\theta$, as in the curve fitting.
All of the films prepared in this study 
had angular distributions that were well fitted by the cosine law; 
however, the films differed with respect to the film material, surface Ra, and molecular orientation.
Specular reflection was not found.

\begin{figure}[t]
	\centering
	\includegraphics[clip, width=0.5\textwidth]{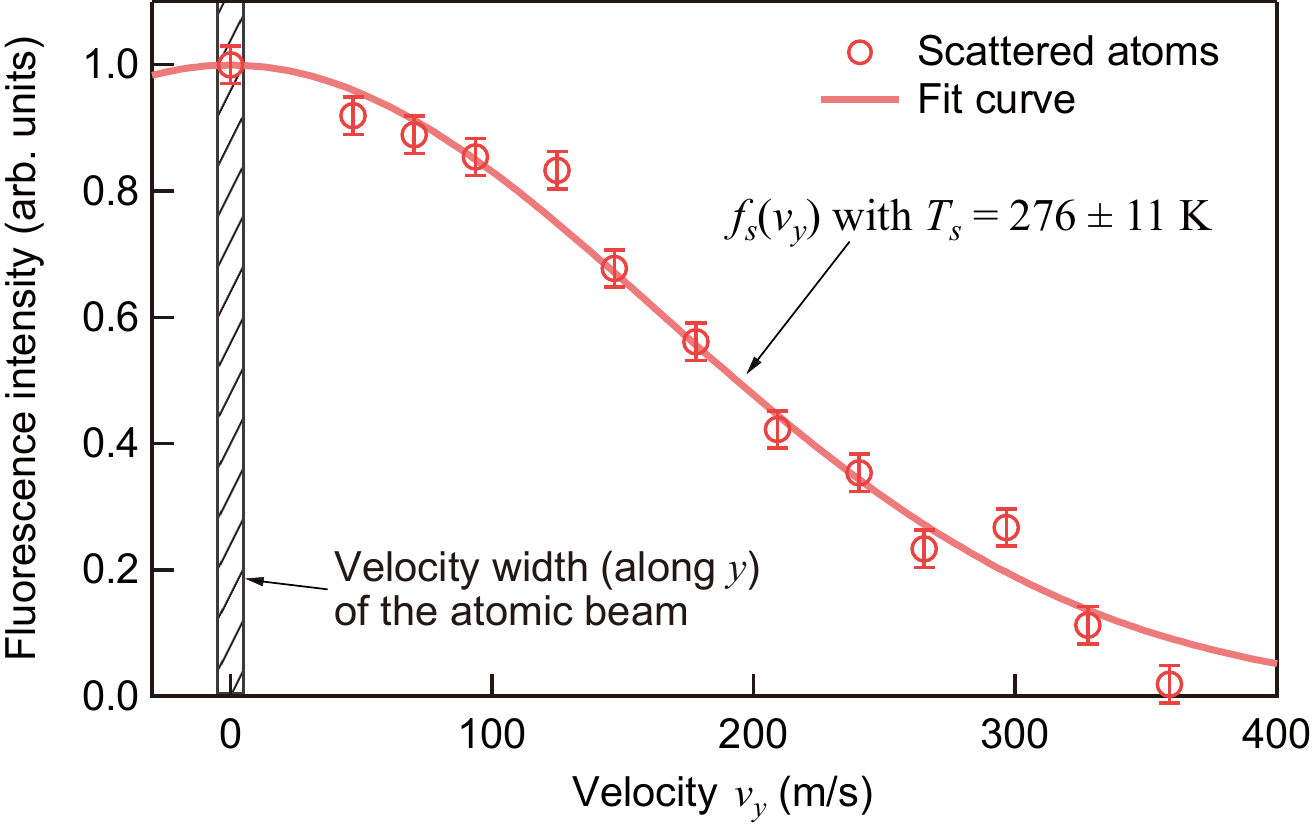}
	\caption{\label{fig: vd}
	Distribution of $v_{y}$ for atoms scattered from film \#2 (red points).
	Horizontal axis shows the selected velocity 
	corresponding to the transition $F=3 \to F'=4$.
	The solid line represents a curve fitted with a Maxwell velocity distribution.
	The hatched area indicates the velocity width along the $y$ axis of the atomic beam.}
\end{figure}

The velocity distribution along the $y$ axis is shown in Fig.~\ref{fig: vd} for film \#2.
The horizontal axis represents the selected velocity $v_{y}$ described by Eq.~(\ref{eq: sv})
with $\Delta_{F'=4} = 0$.
The open circles show the fluorescence intensity and 
the error bars indicate the standard deviation estimated from multiple measurements at a given velocity.
The hatched area indicates the typical velocity width along the $y$ axis of the atomic beam.
It is clear that the velocities of the scattered atoms were distributed over a much broader range than the velocity distribution of the atomic beam.
Curve fitting using Maxwell velocity distribution was considered reasonable, 
given that the angular distributions followed the cosine law, derived directly from Maxwell-Boltzmann statistics. 
The Maxwellian fit curve $f_{s}(v_{y})$ shown by the solid line is given by
\begin{equation}
	f_{s}(v_{y}) = 
	A\exp \left(
	-\frac{mv_{y}^{2}}{2k_{B}T_{s}}
	\right).
\end{equation}
The amplitude $A$ and the temperature $T_{s}$ are data-fitting parameters.

\begin{figure}[t]
	\centering
	\includegraphics[clip, width=0.5\textwidth]{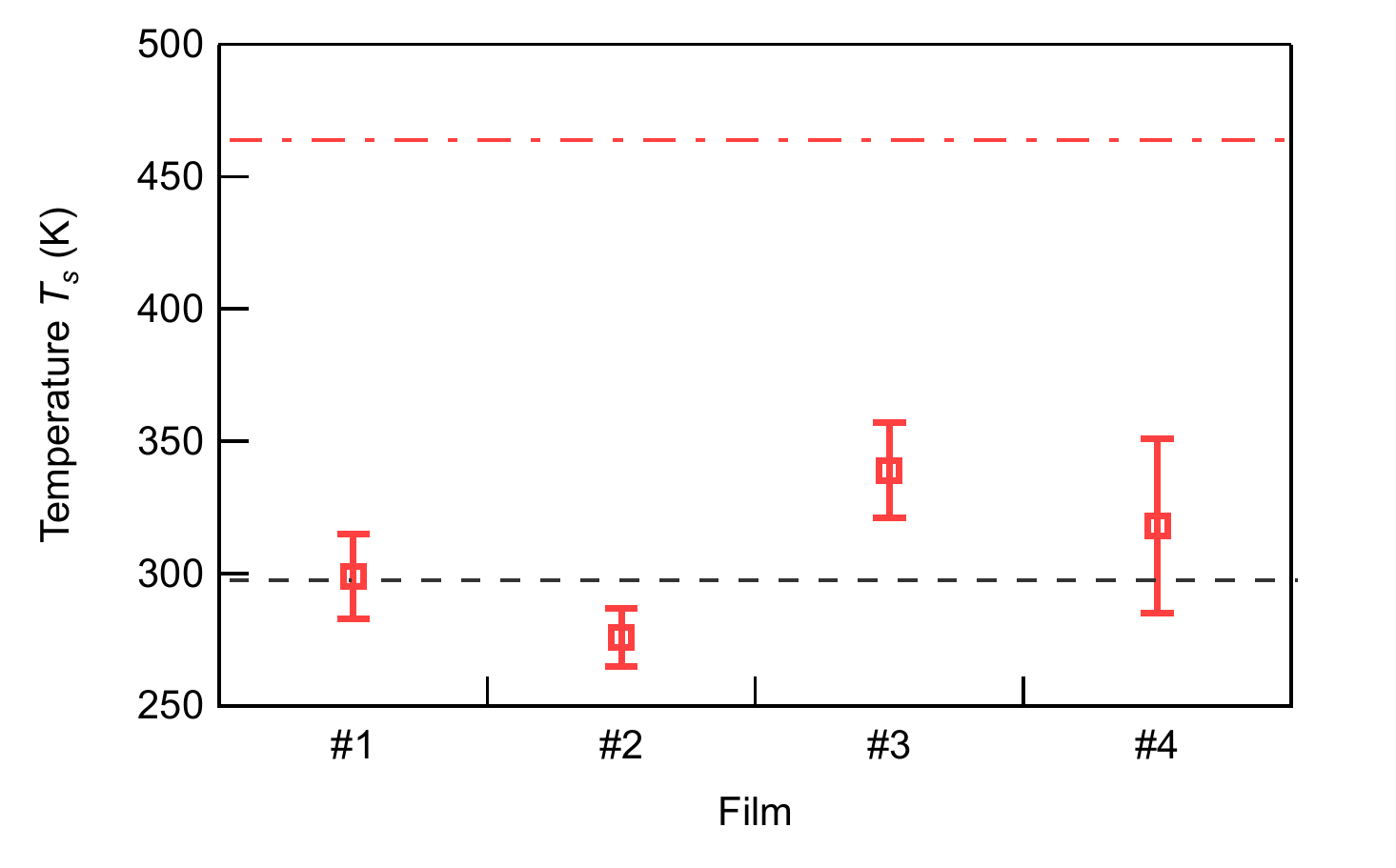}
	\caption{\label{fig: temps} (Color online)
	Temperatures $T_{s}$ derived from fitting the velocity distribution measurements.
	The dashed-dotted line (red) and 
	the dashed line (black) show the temperatures of the atomic beam and the films, 
	respectively.}
\end{figure}

As a result of curve fitting, 
the temperature of the scattered atoms was estimated to be $276 \pm 11$~K for film \#2.
The temperatures of scattered atoms for all films are shown in Fig.~\ref{fig: temps}.
The error bars were obtained from the fitting.
If atoms are reflected elastically by a surface that is sufficiently rough for diffusive reflection, 
the angular distribution corresponds to the cosine law; however, the velocity shows a Maxwellian distribution characterized by the atomic-beam temperature.
In fact, the estimated temperatures were clearly lower than the atomic-beam temperature, as
shown by the dashed-dotted line in Fig.~\ref{fig: temps}.
Moreover, the temperatures were close to the temperature of the film (room temperature), as 
shown by the dashed line.
We therefore concluded, from the cosine-law angular distributions and the temperatures of scattered atoms, that the scattered atoms reached thermal equilibrium with the films.

\section{Conclusions%
\label{sec: conclusions}%
}
We performed scattering experiments of an Rb atomic beam 
on paraffin films.
The paraffin films were prepared using several different procedures.
The surface morphologies and crystal structures of the prepared films were analyzed by AFM and XRD, respectively.
The surface characteristics of the films differed considerably.
By comparing the polarizations of the atomic beam and the scattered atoms, we confirmed that the films preserved hyperfine polarization during the scattering process.
We then measured the angular and velocity distributions of scattered atoms.
Our results indicated that
the cosine law well described the angular distributions of all films, 
despite their different Ras and crystal structures.
The velocity distribution in the direction perpendicular to the incident plane of the atomic beam was fitted by the Maxwell distribution.
The temperatures of the scattered atoms for all films
were much closer to the film temperature than to that of incident atoms.
Based on these results, 
we conclude that 
the incident atoms on the films were well thermalized with the films, and that 
spin polarization was preserved during the scattering process.

To our knowledge, this study is the first to have conducted direct measurements of the angular and velocity distributions of alkali atoms scattered by an anti-spin-relaxation coating.
Accurate representation of these distributions is essential for efficient loading of alkali atoms in miniaturized 
coated-device applications, as well as for research that 
uses short-lived alkali atoms.
Further detailed scattering experiments are expected to provide fundamental information on the interactions between alkali atoms and the coating, for example, dwell time measurement via time-of-flight analysis \cite{Hur79}.

\begin{acknowledgments}
N.S. is a Research Fellow of the Japan Society for the Promotion of Science (JSPS). 
This work was supported by JSPS KAKENHI Grant Numbers JP15H01013, JP17J03089, JP17H02933.
\end{acknowledgments}

\end{document}